\documentstyle[12pt]{article}

\large
\newcommand{\be}{\begin{eqnarray}}
\newcommand{\ee}{\end{eqnarray}}
\newcommand{\bfl}{\begin{flushleft}}
\newcommand{\efl}{\end{flushleft}}
\newcommand\ie {{\it i.e. }}
\newcommand\eg {{\it e.g. }}

\newcommand\rarr{\rightarrow}
\newcommand\half{\frac 1 2 }
\newcommand\noi {\noindent}

\begin{document}

\bibliographystyle{nphys}

\pagestyle{empty}
\vfill
\eject

\vskip 3cm

\vspace* {-25 mm}
\begin{flushright}
USITP-93-5 \\
SUNY-NTG-93-8 \\
March 1993
\end{flushright}
\vskip 1cm
\centerline{\large\bf Two-dimensional QCD at infinite $N_c$ and finite T }
\vskip 4cm
\centerline{\bf{T.H. Hansson$^\dagger$ and   I. Zahed$^\star$} }
\vskip 20mm\noi
\centerline{\bf ABSTRACT}
\vskip 5mm

We analyze two-dimensional large $N_c$ QCD at finite temperature and show
explicitly that the free energy has the correct $N_c$ dependence.

\vfil\noi
$^\dagger$
Department of Physics, Stockholm University \\
Box 6730, S-113 85, Stockholm, Sweden \\
email: hansson@vana.physto.se
\vskip 2mm
\noi$^\star$Nuclear theory group, Department of physics, SUNY at  Stony Brook
\\
Stony Brook, New York, 11794, USA \\
email: zahed@sbnuc1.physics.edu

\vskip 3mm \noindent
${^\dagger}$Supported by the Swedish Natural Science Research Council. \\
${^\star}$Supported in part by the Department of Energy under Grant No.
DE-FG02-88ER40388.
\eject
\newcommand\nc {$N_c$ }

The large \nc expansion provides a qualitative explanation for many strong
interaction phenomena.\cite{witt2} It may well be that the problem of
confinement is nothing more (or rather nothing less) than summing all
planar graphs\cite{gree1}. Also, the early speculations
that these graphs correspond
to an effective string theory has recently received new attention (see \eg
\cite{gros1} and references therein).
The large \nc expansion together with the assumption of confinement implies
that
the number of degrees of freedom (d.o.f.)
at low energy is vastly smaller than expected
from perturbation theory. The low-lying spectrum consist of glueballs and
$q\overline q$ mesons with degeneracy factors independent of $N_c$, while
the perturbative number of d.o.f. are $N_c^2$ for
gluons and $N_c N_f$ for fermions. As discussed by Thorn\cite{thor1} and
Pisarski,\cite{pisa2}
this has important consequences for the finite temperature deconfinement phase
transition. The basic observation is that the free energy in the
low-temperature phase is $O(1)$, while in the high temperature (deconfined)
phase it is $O(N_c)$. If we believe that solving QCD in the large \nc limit
corresponds to summing the planar graphs, we are faced with the following
problem. Why is it that the graphs of fig. 1  do not give a contribution
to the free energy to order $N_c^2$? In this paper we shall address a similar
but much simpler question in the context of the t'Hooft model, \ie large $N_c$
two-dimensional QCD. Here there are no dynamical d.o.f. associated with the
gluo
n
field so the diagrammatic expansion for the partition function always includes
at least one quark line. The free energy is of the form ($\beta = 1/T$),
\be
-T\ln Z = \beta F = N_c F_1(\beta,g^2) + F_0(\beta,g^2) +
\frac 1 {N_c} F_{-1}(\beta,g^2) + ..... \ \ \ \ \ ,
\ee
as illustrated in fig.2. Since the theory confines, and the spectrum is given
by an infinite number of meson states, we expect the leading order term in $F$
to be of order $N_c^0$, that is a free meson gas. The relevant diagrams are
shown on the second line of fig. 2, where it is
also indicated how they can be resummed into meson graphs. Since the mesons
interact weakly with a strength of order $1/\sqrt{N_c}$\cite{witt2} we expect
corrections to $F$ of order $1/N_c$ corresponding to the diagrams on line 3
of fig. 2 and so on.

Using the notation of ref. \cite{thoo1} the full mass-less
planar quark propagator in  light-cone gauge, $A_- = 0$, is
\be
S(k) = \frac {-ik_-} {2k_+k_- - k_- \Sigma(k_-) - i\epsilon}
 \label{prop} \ \ \ \ \ ,
\ee
where the full planar self energy part is
\be
k_-\Sigma(k_-)  = \frac {g^2} \pi \left( 1 - \frac {|k_-|} \lambda \right)
 \label{self}  \ \ \ \ \ .
\ee
Here $\lambda$ is an infrared cutoff that is to be taken to zero at the end of
the calculation. As pointed out in \cite{einh1}, in the limit $ \lambda \rarr
0$, $\lambda$ is nothing but a gauge parameter, so the dispersion relation for
the fermion is explicitly gauge dependent.
It is important to realize that
even though the pole in the propagator moves to infinity in the $\lambda \rarr
0$ limit, this does not constitute a proof of confinement. Rather, since
$\lambda$ is a gauge parameter, one must show that any calculated observable is
independent of $\lambda$ in the $\lambda \rarr 0$ limit. We shall now prove
that for any finite temperature and coupling constant, $F_1(\beta, g^2)$ in
(1), corresponding to the diagrams in the first line in fig.2,
is identically zero in the limit $\lambda \rarr 0$.

The main idea is to relate the leading piece, ${\cal F}_1 = F_1(\beta, g^2)/L$
($L$ is the volume), in the free energy density to the propagator via the
relation,
\be
g^2 \frac {d^2} {dg^2} {\cal F}_1 = \half {\rm Tr} (S\Sigma )  \label{relation}
\ee
which we now prove using diagrammatic techniques.

Consider fig. 3a which shows all diagrams contributing to ${\cal F}_1$, and
where the shadowed blob represent all the ways the n gluon lines can connect to
each other to form a planar graph.
(Note that there are no planar vertex corrections.)
We have explicitly shown the power of $g^2$
and there is a crucial statistical weight factor $1/2n$ due to the 2n-fold
symmetry of the diagram. In fig. 3b we show a similar set of
diagrams contributing to the difference between the full and the free
propagator - note the absence of the statistical weight factor. From the figure
we immediately get,
\be
g^2 \frac {d^2} {dg^2} {\cal F}_1 = \half {\rm Tr}
\left[ (S-S_0)S_0^{-1}\right]  \ \ \ \ \ ,
\ee
and substituting the Dyson equation $S = S_0 + S\Sigma S_0$ we get
(\ref{relation}).
This relation can also be derived by noticing that in the
light cone gauge the gluonic action is quadratic in $A_+$, and performing this
Gaussian integral we get the following induced four Fermi interaction
\be
{\cal L} = \frac {g^2}2 \int d^2x\,d^2y \,D(x-y)\,
                  \overline{\psi}\lambda^a\gamma^-\psi (x) \,
                 \overline{\psi}\lambda^a\gamma^-\psi (y)
    \label{interaction} \ \ \ \ \ .
\ee
Now, the free energy can be expressed as,
\be
\frac {d^2} {dg^2} {\cal F} =
\frac 1 2 \int d^2x\,d^2y D(x-y)
                 \langle  \overline{\psi}\lambda^a\gamma^-\psi (x)\,
                 \overline{\psi}\lambda^a\gamma^-\psi (y) \rangle
      \label{nonlocal} \ \ \ \ \ ,
\ee
where $D(x-y)$ is the gluon propagator, and the expectation value
is with respect to the induced (non-local)
four Fermi interaction (\ref{interaction}).
In the large $N_c$ limit (\ref{nonlocal}) reduces to (\ref{relation})
using the Dyson equation for $\Sigma$ given in \cite{thoo1}.

Before substituting (\ref{prop}) and (\ref{self}) in (\ref{relation}) we
notice that the pole in (\ref{prop}) is given by,
\be
E_p = -k + \frac { g^2} {\sqrt{2}\pi\lambda} {\rm sgn}(E-k)
\label{pole} \ \ \ \ \ ,
\ee
where we have reintroduced the usual energy and momentum via $\sqrt 2 k_\pm = E
\pm k$. This dispersion relation, which is illustrated in fig. 4, will change
at finite temperature, but in the limit $\lambda \rarr 0$, where the mass gap
diverges, any finite temperature will not influence the spectrum.
Thus, the only place where the temperature enters the calculation, is in the
loop integration implied by the trace in the RHS of eq. (\ref{relation}). We
get,
\be
\frac {d^2} {dg^2} \left( {\cal F}_1  - {\cal F}^{vac}_1\right) =
\frac 1 {2\pi}\int \frac {dk} {2\pi}\, \int_{-i\infty}^{i\infty}
\frac {dE} {2\pi i} \, n_\beta(E)  Ev \left(
\frac {1 - |k_-|/ \lambda } {2k_+k_- - k_- \Sigma(k_-) - i\epsilon} \right)
    \label{integral} \ \ \ \ \ ,
\ee
where $n_\beta(E)$ is the Fermi distribution function and $Ev f(E) = f(E) +
f(-E)$. The temperature independent vacuum part, ${\cal F}^{vac}_1$,
is of no concern here.
Doing the $E$ integration by contours, we pick up the pole at
$E_p$ given by (\ref{pole})
and because of the mass- gap, the integral (\ref{integral}) is suppressed by
a factor $n_\beta(E_p) \sim e^{-\beta/\lambda}$. Thus for any fixed
temperature, the integrand goes to zero in the limit $\lambda \rarr 0$ and thus
the whole integral equals zero. We have now proved that there is no
$g$-dependence in the $O(N_c)$ contribution to the free energy. We cannot
logically exclude the possibility of a $g^2$ independent piece $cT^2$,
where $c$ is an $g$ (and $\lambda$) independent constant.
Such a behaviour is, however, expected for a gas of free, mass-less particles,
and we find it extremely unlikely that such a contribution should be present in
the $g \rightarrow\infty$ limit. We thus conclude that the $O(N_c)$
contribution
to the free energy is identically zero for all $g$.

Our argument breaks down at infinite temperature (or equivalently at
vanishing $g$, since what matters is the dimensionless quantity $\beta g$)
where
$T\sim 1/\lambda\sim\infty$. In this regime, the thermal effects can overcome
the string tension $\sigma\sim g^2$ allowing for a possible
phase change. This can also be made plausible by recalling
that the number of mesons in two dimensions
grow quadratically with the energy ($n\sim T^2/\sigma$). As a result, the
free energy at high temperature is of order $nT\sim T^3/\sigma$. A phase change
is expected when the latter is of order $N_fN_cT$, $\,\,i.e.$
$ T_c\sim g\sqrt{N_c}\rightarrow\infty$.\footnote{
Of course, at finite \nc there cannot be any finite $T$ phase transition. What
happens at infinite \nc is a tricky question and this argument is only
suggestive.}
Finally,
if we think of $\lambda$ as roughly the size of the system, we conclude that
the large \nc limit does not commute with the thermodynamical limit (\ie
taking the box size to infinity at fixed $T$).

We believe that this result is illustrative to what happens in four
dimensions, where we expect the low temperature limit of QCD to be dominated
by pions. The strong infrared divergences in the quark and gluon sector at
low energy will cause the order $N_c^2$ (gluonic) and $N_cN_f$ (fermionic)
terms in the free energy to vanish identically, just as in 2-dimensional QCD.
In four dimensions, however, we expect a phase transition to
take place at $T_c \sim \Lambda$. As we approach this temperature from below,
the mesonic interactions, although suppressed by $1/N_c$, will be dominant
since the number of particles is likely to grow exponentially in a Hagedorn
manner. Above the critical temperature, the quark - gluon description is
believed to be more economical even though high temperature perturbative QCD is
still plagued by infrared problems.

\vglue .5cm
{\bf \noindent  Acknowledgements } \\
\noi
This work was partially supported by the US DOE grant DE-FG-88ER40388.


\newpage

\vskip 3cm
\noindent{\bf Figure Captions}
\vskip 2cm
\noindent Fig. 1 : Leading order contribution to the free energy in QCD.
\vskip .3cm
\noindent Fig. 2 : Contribution to the free energy in two dimensional QCD.
\vskip .3cm
\noindent Fig. 3 : (a) Diagrams contribution to the leading part of the free
 energy where $g^2$ and statistical factors are shown explicitly. \\
(b) Corresponding
graphs for the difference between the full planar propagator and the free one.
\vskip .3cm
\noindent Fig. 4 : The quark dispersion relation in light-cone gauge.

\end{document}